\documentclass[a4paper]{jpconf}
\usepackage{graphicx}
\begin{document}
\title{Emergence of Yang Mills theory from the Non-Abelian Nambu Model}

\author{C. A. Escobar$^{1}$ and L. F. Urrutia$^{1}$}

\address{$^{1}$ Instituto de Ciencias Nucleares, Universidad Nacional Aut{\'o}noma de
M{\'e}xico, A. Postal 70-543, 04510 M{\'e}xico D.F., M{\'e}xico}

\ead{carlos.escobar@correo.nucleares.unam.mx}

\begin{abstract}

The equivalence between the Non-Abelian Nambu model (NANM) and Yang Mills theory is proved, after demanding the Gauss laws at some initial time to the first one. Thereby, the Lorentz violation encoded into the constraint that defines the NANM is physically unobservable.  As result, the Goldstone bosons in the NANM arising from the spontaneous symmetry breaking can be identified as the standard gauge fields.

\end{abstract}


Lorentz invariance violation (LIV) has attracted considerable attention since some theories, such as Quantum Gravity \cite{quantum1} and Strings \cite{String2}, provide some mechanisms where the fundamental Lorentz symmetry can be broken.  Some schemas seek for deviations of Lorentz symmetry experimentally measurable \cite{SME}, while others attempt to incorporate the concept of LIV in such a way that the physical effects are unobservable. As an example of the last idea, the Nambu model \cite{Nambu} is motivated by means of Lorentz violation arising from a spontaneous symmetry breaking. Perturbative and nonperturbative calculations show that, under some conditions, the Nambu model is equivalent to the standard Electrodynamics \cite{Abelian}. In this way, the Nambu model provides a dynamical interpretation of the corresponding massless gauge bosons in terms of the Goldstone bosons (GB).  Additional previous works in Yang Mills theories and gravitation can be found in Refs. \cite{YM2} and Refs. \cite{Grav}, respectively.

In the present work, following a Hamiltonian analysis we show a nonperturbative equivalence between the Non-Abelian Nambu model and Yang Mills (YM) theory. To prove the foregoing equivalence  we have to deal with the issues: i) the number of degrees of freedom (DOF) in the NANM is larger than that of the Yang Mills theory, ii) the NANM is not gauge invariant, iii) the equations of motion of the NANM and Yang Mills theory do not match and iv) current conservation does not hold in the NANM. The equivalence will be proved showing that by means of a suitable canonical transformation of the NANM variables plus the Gauss laws as initial conditions, both the Hamiltonian and canonical algebra of the NANM are the same as those describing the Yang Mills theory. The appendix includes a review of the Hamiltonian analysis of Yang Mills theory, which we use as a benchmark to state the proposed equivalence.

\section{The Non-Abelian Nambu model}

The Non-Abelian Nambu model is defined by the Lagrange density
\begin{equation}
\mathcal{L}(A_{\mu }^{a})=-\frac{1}{4}F_{\mu \nu }^{a}F^{a\mu \nu }-J^{a\mu
}A_{\mu }^{a},  \label{YML}
\end{equation}
plus the condition
\begin{equation}
A_{\mu }^{a}A^{a\mu }=n^{2}M^{2},\qquad M^{2}>0,\quad \mu =0,1,2,3,\quad
a=1,...,N.  \label{NANMC}
\end{equation}
Here, $n_\mu$ is a properly oriented constant vector, such that $n^2=\pm1,0$, while $M$ is the proposed scale associated with the spontaneous Lorentz symmetry breaking (SLSB) and $\mu,\nu=0,1,2,3$ and $a=1,2,...,N$ indices in the Lorentz and gauge group with $N$ generators, respectively. The constraint (\ref{NANMC}) can be understood as providing a nonzero vacuum expectation value $\langle A_{\mu }\rangle=n_{\mu }M$, which produces spontaneous symmetry breaking of Lorentz invariance and the appearing of Goldstone bosons via the Goldstone theorem.

A convenient way to deal with the different cases of the NANM is by means of the parametrization 
\begin{equation}
A_{0}^{a}=B^{a}\left( 1+\frac{N}{4B^{2}}\right),\,\,\,\,\,\,\,\,\,\,\,\,\,\,\,
A_{3}^{a}=B^{a}\left( 1-\frac{N}{4B^{2}}\right), 
\label{GEN_PARAM}
\end{equation}
with
\begin{equation}
N=\left( A_{\bar{\imath}}^{b}A_{\bar{\imath}}^{b}+n^{2}M^{2}\right) ,\;\;\;
4B^{2}\pm N\neq 0,\quad\quad  
\bar{\imath}=1,2,
\end{equation}
which certainly satisfies the condition (\ref{NANMC}) and it is written in terms of the $3N$ independent GB \ $B^{a},A_{\bar{\imath}}^{b}$. Notice that $n^2$ can take the values $\pm1$ and $0$.

After the substitution of (\ref{GEN_PARAM}) in the Lagrangian density (\ref{YML}), the variation of the corresponding action with respect to $B^{a},A_{\bar{\imath}}^{b}$ yields the equations of motion
\begin{equation}
\delta A_{\bar{\imath}}^{a}:\;\;\;\;\mathcal{E}^{{\bar{\imath}}a}+\frac{B^{b}
}{2B^{2}}\left[ \mathcal{E}^{0b}-\mathcal{E}^{3b}\right] A_{\bar{\imath}
}^{a}=0,  \label{ecumov2}
\end{equation}
\begin{equation}
\delta B^{a}:\;\;\left( \left( 1+\frac{N}{4B^{2}}\right) \delta ^{ab}-
\frac{N}{4B^{2}}\frac{2B^{b}B^{a}}{B^{2}}\right) \mathcal{E}^{0b}+\left(
\left( 1-\frac{N}{4B^{2}}\right) \delta ^{ab}+\frac{N}{4B^{2}}\frac{
2B^{b}B^{a}}{B^{2}}\right) \mathcal{E}^{3b}=0,  \label{ecumov3}
\end{equation}
with the notation
\begin{equation}
\mathcal{E}^{\nu a}=\left( D_{\mu }F^{\mu \nu }-J^{\nu }\right)^{a}.
\label{EQMOTYM}
\end{equation}

Let us recall that in the case of the $SO(N)$ Yang Mills theory the equations of motion are just given by $\mathcal{E}^{\nu a}=0$. The equations of motion (\ref{ecumov2}) and (\ref{ecumov3}) do not imply current conservation $D_{\nu }J^{\nu a}=0$, basically because the condition (\ref{NANMC}) breaks the gauge invariance. A way to recover the Yang Mills equations of motion together with gauge invariance is to impose the Gauss laws $\mathcal{E}^{0a}=0$. In this way, under the conditions $4B^2\pm N \neq 0$, Eq. (\ref{ecumov3}) yields the solution $\mathcal{E}^{3b}=0$. These two conditions in (\ref{ecumov2}) provide the final set $\mathcal{E}^{{\bar{\imath}}a}=0.$

In order to unify the notation when going to the Hamiltonian formulation we
introduce the $3N$ DOF $\Phi _{A}^{a},\;A=1,2,3,$ 
\begin{equation}
\Phi _{1}^{a}=A_{1}^{a},\;\;\;\;\Phi _{2}^{a}=A_{2}^{a}\;,\;\;\;\Phi
_{3}^{a}=B^{a},\;
\end{equation}
in such a way that the coordinate transformation 
\begin{equation}
A_{i}^{a}=A_{i}^{a}(\Phi _{A}^{b}),  \label{NACHVAR}
\end{equation}
arising from (\ref{GEN_PARAM}) is invertible. The relevant property of the transformation (\ref{NACHVAR}) is
that 
\begin{equation}
\dot{A}_{i}^{a}=\frac{\partial A_{i}^{a}}{\partial \Phi _{B}^{b}}\dot{\Phi}
_{B}^{b},\;\;\;\;\rightarrow \;\;\;\frac{\partial \dot{A}_{i}^{a}}{\partial 
\dot{\Phi}_{B}^{b}}=\frac{\partial A_{i}^{a}}{\partial \Phi _{B}^{b}},
\label{FUNDCT}
\end{equation}
together with the invertibility of the velocities 
\begin{equation}
\dot{\Phi}_{A}^{a}=\frac{\partial \Phi _{A}^{a}}{\partial A_{i}^{b}}\dot{A}
_{i}^{b}.  \label{INVERTVEL}
\end{equation}
In the following we will not require the explicit form of the transformations (\ref{GEN_PARAM}), but only its generic form (\ref{NACHVAR}), together with the property that this transformation can be inverted.

Next we proceed to calculate the Hamiltonian density of the NANM in terms of the canonically conjugated variables $\Phi _{A}^{b},\;\Pi _{A}^{b}$ and employing a procedure that allows to make direct contact with both the YM Hamiltonian density (\ref{hamiltonian_standar}) and the YM canonical algebra (\ref{FINALDB}). After the substitutions (\ref{NACHVAR}) together with $A_{0}^{a}=A_{0}^{a}(\Phi _{A}^{b})$ are made, the Lagrangian density (\ref{YML}) can be splitted as 
\begin{equation}
\mathcal{L}_{\rm NANM}(\Phi ,\dot{\Phi})=\frac{1}{2}E_{i}^{a}E_{i}^{a}-\frac{1}{2
}B_{i}^{a}B_{i}^{a}-J^{a\mu }A_{\mu }^{a},\,
\end{equation}
where 
\begin{equation}
E_{i}^{a}=\dot{A}_{i}^{a}-D_{i}A_{0}^{a},\;\;\;\;\;\;\;B_{i}^{a}=\frac{1}{2}
\epsilon _{ijk}F_{jk}^{a},\;  \label{COLOREB}
\end{equation}
with $E_{i}^{a}=E_{i}^{a}(\Phi ,\dot{\Phi}),\,B_{i}^{a}=B_{i}^{a}(\Phi )$. The canonically conjugated momenta are calculated as 
\[
\Pi _{A}^{a}=\frac{\partial \mathcal{L}_{\rm NANM}(\Phi ,\dot{\Phi})}{\partial 
\dot{\Phi}_{A}^{a}}=E_{i}^{b}\frac{\partial \dot{A}_{i}^{b}}{\partial \dot{
\Phi}_{A}^{a}}=E_{i}^{b}\frac{\partial A_{i}^{b}}{\partial \Phi _{A}^{a}}, 
\]
according to (\ref{FUNDCT}). The inverse of Eqs. (\ref{NACHVAR}) allows us to write the colored electric fields $E_{i}^{a}$ as functions of the momenta $\Pi _{A}^{b}$ of the NANM
\begin{equation}
E_{i}^{b}(\Phi ,\Pi )=\frac{\partial \Phi _{A}^{a}}{\partial A_{i}^{b}}\Pi
_{A}^{a}.  \label{MOMCT}
\end{equation}
It can be proved that the Wronskian of the system is
\begin{equation}
\det \left( \frac{\partial ^{2}\mathcal{L}_{\rm NANM}(\Phi ,\dot{\Phi})}{
\partial \dot{\Phi}_{A}^{a}\partial \dot{\Phi}_{B}^{b}}\right) =\det \left( 
\frac{\partial \Pi _{A}^{a}}{\partial \dot{\Phi}_{B}^{b}}\right) =\det
\left( \frac{\partial \dot{A}_{i}^{c}}{\partial \dot{\Phi}_{B}^{b}}\frac{
\partial A_{i}^{c}}{\partial \Phi _{A}^{a}}\right) =\det \left( \frac{
\partial A_{i}^{c}}{\partial \Phi _{B}^{b}}\frac{\partial A_{i}^{c}}{
\partial \Phi _{A}^{a}}\right) \neq 0.
\end{equation}
In this way, the NANM is exhibited as a regular system in the parameterization (\ref{GEN_PARAM}), so that no constraints are present.

The NANM Hamiltonian density is 
\begin{equation}
\mathcal{H}_{\rm NANM}=\;\Pi _{A}^{a}\dot{\Phi}_{A}^{a}-\left( \frac{1}{2}
E_{i}^{a}E_{i}^{a}-\frac{1}{2}B_{i}^{a}B_{i}^{a}-J^{a\mu }A_{\mu
}^{a}\right) ,
\end{equation}
which we rewrite in successive steps as
\begin{eqnarray}
\mathcal{H}_{\rm NANM}(\Phi ,\Pi ) &=&\;\frac{1}{2}E_{i}^{a}E_{i}^{a}+\frac{1}{2}
B_{i}^{a}B_{i}^{a}-\left( D_{i}E_{i}^{b}-J^{b0}\right)
A_{0}^{b}+J^{ai}A_{i}^{a},  \label{HNANM_FIN}
\end{eqnarray}
where we have used the relations (\ref{INVERTVEL}), (\ref{COLOREB}) and (\ref{MOMCT}), together with an integration by parts in the term containing the covariant derivative. The dependence of $\mathcal{H}_{\rm NANM}$ upon the canonical variables $\Phi ,\Pi$ is clearly established by the change of variables (\ref{NACHVAR}) and (\ref{MOMCT}). The NANM canonical variables satisfy the standard PB algebra
\begin{equation}
\left\{ \Phi _{A}^{a}(\mathbf{x}),\Phi _{B}^{b}(\mathbf{y})\right\}
=0,\;\;\;\left\{ \Pi _{A}^{a}(\mathbf{x}),\Pi _{B}^{b}(\mathbf{y})\right\}
=0,\;\;\;\;\left\{ \Phi _{A}^{a}(\mathbf{x}),\Pi _{B}^{b}(\mathbf{y}
)\right\} =\delta ^{ab}\delta _{AB}\delta ^{3}(\mathbf{x-y}).  \label{NANMPB}
\end{equation}
Now we can consider the NANM Hamiltonian density\ (\ref{HNANM_FIN}) from the perspective of the fields $A_{i}^{a},\;E_{i}^{a}$. The following relation arising from the velocity dependent term of NANM\ Hamiltonian action
\begin{equation}
\int d^{4}x\;\Pi _{A}^{a}\dot{\Phi}_{A}^{a}=\int d^{4}x\;E_{i}^{a}\dot{A}
_{i}^{a}= \int d^{4}x\;(-E^{ai})\dot{A}_{i}^{a},
\end{equation}
establishes $(-E^{ai})$ as the canonically conjugated momenta of $A_{i}^{a}$. In this way (\ref{HNANM_FIN}) can be read as a Hamiltonian density $\mathcal{H}(A,E)\;$ obtained from $\mathcal{H}_{NANM}(\Phi ,\Pi )$ via the substitution of the phase space transformations 
\begin{equation}
(\Phi ,\Pi )\rightarrow (A,E),  \label{PSTRANS}
\end{equation}
which follow from the inverses of Eqs.(\ref{NACHVAR}) and (\ref{MOMCT}) plus the relation 
\begin{equation}
A_{0}^{a}=\frac{A_{3}^{a}}{\sqrt{A_{3}^{b}A_{3}^{b}}}\left( \sqrt{
A_{i}^{b}A_{i}^{b}+n^{2}M^{2}}\right) ,  \label{A0INV}
\end{equation}
in terms of the new variables. But, since the transformations (\ref{MOMCT}) are generated by the change of variables (\ref{NACHVAR}) in the coordinate space, we know from classical mechanics that the full transformation in the phase space is a canonical transformation. In this way we automatically recover the PB algebra
\begin{equation}
\left\{ A_{i}^{a}(\mathbf{x}),A_{j}^{b}(\mathbf{y})\right\} =0,\;\;\;\left\{
E^{ai}(\mathbf{x}),E^{bj}(\mathbf{y})\right\}=0,\;\;\;\;\left\{ A_{i}^{a}(
\mathbf{x}),E^{bj}(\mathbf{y})\right\} =-\delta ^{ab}\delta
_{i}^{j}\delta^{3}(\mathbf{x-y})  \label{AEPBALG}
\end{equation}
from Eq. (\ref{NANMPB}). Summarizing, from each Hamiltonian version of the NANM, defined by the different values of $n^{2}$, we can regain, via a canonical transformation, the Hamiltonian density (\ref{HNANM_FIN}) together with the canonical algebra (\ref{FINALDB}). The Hamiltonian density (\ref{HNANM_FIN}) differs from the YM Hamiltonian density (\ref{hamiltonian_standar}) only in the fact that the Gauss laws $\Omega^{b}=\left( D_{i}E_{i}^{b}-J^{b0}\right) =0$ do not appear as first class constraints, because $A_{0}^{a}$ are not arbitrary Lagrange multipliers, but functions of the coordinates as shown in Eq. (\ref{A0INV}). To deal with this issue we study the time evolution of the Gauss functions under the dynamics of the NANM, starting from the Hamiltonian density (\ref{HNANM_FIN}). After some calculations, in the NANM we find \cite{Conditions}

\begin{equation}
\dot{\Omega}^{a}=-gC^{abc}A_{0}^{b}\Omega ^{c}-D_{\mu }J^{\mu a}+D_{3}\left( 
\frac{A_{0}^{1}}{A_{3}^{1}}\Omega ^{a}\right) -D_{3}\left( \frac{A_{3}^{a}}{
A_{3}^{c}A_{3}^{c}}A_{0}^{b}\Omega ^{b}\right) +D_{i}\left( \frac{A_{i}^{a}}{
N}A_{0}^{b}\Omega ^{b}\right),
\label{DervivGauss}
\end{equation}
where $A_{0}^{b}\;$is given by Eq. (\ref{A0INV}). It follows that, imposing the Gauss constraints as initial conditions ($\Omega^a(t=t_0)=0$) upon Eqs. (\ref{ecumov2})-(\ref{ecumov3}), the standard Yang-Mills equations of motion ($\mathcal{E}^{\nu a}=0$) are recovered and they are valid at $t=t_0$. As a consequence of the antisymmetry of Maxwell tensor, the relation 
\begin{equation}
0=D_\nu \mathcal{E}^{\nu a}= D_\nu (D_\mu F^{\mu\nu}-J^\nu)^a= -D_\nu J^{\nu a}=0, \quad \quad (\textrm{at } t=t_0),
\end{equation}
holds, yielding current conservation at $t=t_0$. Using $\Omega^a(t=t_0)=0$ and $D_\nu J^{\nu a}|_{t=t_0}=0$ in Eq. (\ref{DervivGauss}), we obtain $\dot{\Omega}^a(t=t_0)=0$. Since the relations $\Omega^a(t=t_0)=0$ and $\dot{\Omega}^a(t=t_0)=0$ are fulfilled, we obtain
\begin{eqnarray}
\nonumber
\Omega^a(t=t_0+\delta t_1)&=&\Omega^a(t=t_0)+\dot{\Omega}^a(t=t_0)\delta t_1+\cdots, \\
&=&0.
\end{eqnarray}
Using that $\Omega^a(t=t_0+\delta t_1)=0$, we observe that the Yang Mills equations are now valid at $t=t_0+\delta t_1$, which again, due to antisymmetry of the Maxwell tensor, imply current conservation at $t=t_0+\delta t_1$ and $\dot{\Omega}^a(t=t_0+\delta t_1)=0$ via Eq. (\ref{DervivGauss}).
The relations $\Omega^a(t=t_0+\delta t_1)=0$ and $\dot{\Omega}^a(t=t_0+\delta t_1)=0$ imply $\Omega^a(t=t_0+\delta t_1+\delta t_2)=0$. Iterating the previous process, it follows that the Gauss laws and current conservation are now valid for all time. Summarizing, the above analysis shows that in the case of the  Non-Abelian Nambu model, the specific gauge structure of the theory allows us to  impose only the  Gauss  constraints as initial conditions, which necessarily yield current conservation for all time. 
 In this way we can recover the $SO(N)$ Yang-Mills theory by imposing the Gauss laws as Hamiltonian constraints, with arbitrary functions $N^{a}$ adding $-N^{a}\Omega ^{a}$ to $\mathcal{H}_{\rm NANM}$ and redefining $A_{0}^{a}+N^{a}=\Theta ^{a}$. This leads to 
\begin{equation}
\mathcal{H}_{\rm NANM}=\frac{1}{2}(\mathbf{E}^{2}+\mathbf{B}^{2})-\Theta ^{a}\Omega^{a}+J_{i}^{a}A^{ia},  
\end{equation}
where $\Theta ^{a}$ are now arbitrary functions, thus getting back to the YM Hamiltonian density (\ref{hamiltonian_standar}) plus the canonical algebra (\ref{FINALDB}). It is worth to mention that the imposition of the $N$ Gauss laws as first class constraints in the NANM Hamiltonian reduces the $3N$ degrees of freedom of the NANM to $2N$, which are the number of degrees of freedom of the Yang Mills theory.


\section{Summary}

A nonperturbative equivalence between the Yang Mills theory and the Non-Abelian Nambu model (NANM) has been established, after the Gauss laws are imposed as initial conditions for the latter. The possible interpretation of gauge particles (i.e. photons and gravitons, for example) as the Goldstone bosons (GB) modes arising from some spontaneous symmetry breaking is an interesting hypothesis which would provide a dynamical setting to the gauge principle.
In this work we have taken the Nambu approach in which the spontaneous Lorentz symmetry breaking is incorporated in an effective way in the model, by means of a non-linear constraint. The challenge posed by this setting is to show under which conditions, the violations of Lorentz symmetry and gauge invariance, introduced by the non-linear constraint, are unobservable in such a way that the appearing Goldstone bosons can be interpreted as the gauge particles of an unbroken gauge theory. In the case of the NANM we have proved that such conditions are the Gauss laws just as initial conditions.  The strategy employed to prove the equivalence between both theories was to show that, by means of a suitable transformation of the NANM variables plus the Gauss laws as initial conditions, the Hamiltonian and canonical algebra of the NANM are reduced to their corresponding in the Yang Mills theory, in such a way that both theories are undistinguishable.


\ack
L. F. U has been partially  supported by the project CONACyT \# 237503. C. A. E. and L. F. U acknowledge support from the project UNAM (Direcci\'on General de Asuntos del Personal Acad\'emico) \# IN104815.


\appendix
\setcounter{section}{1}
\section*{Appendix A: The SO(N) Yang-Mills (YM) theory}
\label{APPA}
We present a brief review of the Hamiltonian formulation of the standard $SO(N)$ Yang-Mills theory. The Yang- Mills Lagrangian density is given by 
\begin{equation}
\mathcal{L}=Tr\left[ -\frac{1}{4}\mathbf{F}_{\mu \nu }\mathbf{F}^{\mu \nu }-
\mathbf{J}^{\mu }\mathbf{A}_{\mu }\right] ,  \label{LDENS}
\end{equation}
where boldfaced quantities denote matrices in the Lie algebra of the internal symmetry group $SO(N)$ with $N(N-1)/2$ generators $t^{a}$; i.e. $\mathbf{M}=M^{a}t^{a}$. This algebra is generated by $\;\left[ t^{a},t^{b}\right] =C^{abc}t^{c}$, where the structure constants $C^{abc}$ are completely antisymmetric.

The canonical Hamiltonian density arising from the Lagrange density (\ref{LDENS}) is given by
\begin{equation}
\mathcal{H}=\Pi _{a}^{i}\dot{A}_{i}^{a}-\mathcal{L}=\frac{1}{2}(\mathbf{E}
^{2}+\mathbf{B}^{2})-A_{0}^{a}(D_{i}E_{i}-J^{0})^{a}-J_{i}^{a}A_{i}^{a}.
\end{equation}
We employ the Dirac's method to construct the canonical theory due to the fact that primary constraints 
\begin{equation}
\Sigma ^{a}=\Pi _{0}^{a}\simeq 0,  \label{YMC1}
\end{equation}
are present. The extended Hamiltonian density is
\begin{equation}
\mathcal{H}_{E}=\frac{1}{2}(\mathbf{E}^{2}+\mathbf{B}
^{2})-A_{0}^{a}(D_{i}E_{i}-J^{0})_{a}+J_{i}^{a}A_{a}^{i}+\lambda ^{a}\Sigma
_{a},
\end{equation}
where $\lambda ^{a}$ are arbitrary functions. The evolution condition of the primary constraints $\dot{\Sigma}^{a}(\mathbf{x})=\{\Sigma^a,\mathcal{H}_E \}$ leads to the Gauss laws 
\begin{equation}
\Omega ^{a}=(D_{i}E_{i}-J^{0})^{a}\simeq 0.  \label{YMC2}
\end{equation}
It is not difficult to prove that\ (\ref{YMC1}) \ and (\ref{YMC2}) are the only constraints present and that they constitute a first class set. In fact, calculating the time evolution of $\Omega ^{a}$ yields
\begin{equation}
\dot{\Omega}^{a}=-\partial
_{0}J_{0}^{a}-C^{abc}A_{0}^{b}J_{0}^{c}-C^{abc}A_{0}^{b}\Omega
^{c}-D_{k}J^{ak}=-C^{abc}A_{0}^{b}\Omega ^{c}-D_{\mu }J^{a\mu },
\end{equation}
which is zero, modulo the constraints and using current conservation.

Normally one fixes 
\begin{equation}
\Pi _{0}^{a}\simeq 0,\,\,\,\,\,\,\,\,\,A_{0}^{a}\simeq \Theta^a ,
\label{cond_YM}
\end{equation}
with $\Theta^{a}$ being arbitrary functions to be consistently determined after the remaining first class constraints $\Omega ^{a}\;$are fixed. $\;$

The final Hamiltonian density is
\begin{equation}
\mathcal{H}_{E}=\frac{1}{2}(\mathbf{E}^{2}+\mathbf{B}^{2})-\Theta
^{a}(D_{i}E_{i}-J^{0})_{a}+J_{i}^{a}A_{a}^{i}.  \label{hamiltonian_standar}
\end{equation}
Once $\Pi _{0}^{a}\,\;$and $\,A_{0}^{a}\;$are fixed strongly, the Dirac brackets among the remaining variables are 
\begin{equation}
\{A_{i}^{a}(\mathbf{x},t),A_{j}^{b}(\mathbf{y},t)\}^{\ast
}=0,\,\,\,\,\,\,\,\,\,\,\,\,\{E^{ai}(\mathbf{x},t),E^{bj}(\mathbf{y}
,t)\}^{\ast }=0,\;\;\;\;\{A_{i}^{a}(\mathbf{x},t),E^{bj}(\mathbf{y}
,t)\}^{\ast }=-\delta _{i}^{j}\delta ^{ab}\delta (\mathbf{x}-\mathbf{y}).
\label{FINALDB}
\end{equation}
The final count of degrees of freedom (DOF) per point in coordinate space
yields
\begin{equation}
\#DOF=\frac{1}{2}(2\times 4N-2\times 2N)=2N.
\end{equation}

At this point, the dynamics of Yangs Mills is determined by the Hamiltonian (\ref{hamiltonian_standar}) together with the canonical algebra (\ref{FINALDB}). The goal to obtain the proposed equivalence, between YM theory and the NANM, will be to obtain the same Hamiltonian and canonical algebra from the NANM, after the Gauss laws are imposed as initial conditions.



\section*{References}
\begin{thebibliography}{9}

\bibitem{quantum1} Gambini R and Pullin J 1999 \textit{Phys. Rev} D \textbf{59} 124021; Alfaro J, Morales-T\'ecotl H A and Urrutia L F 2002 \textit{Phys. Rev} D \textbf{65} 103509; Alfaro J, Morales-T\'ecotl H A and Urrutia L F 2000 \textit{Phys. Rev. Lett.} \textbf{84} 2318; Carroll S M, Harvey J A, Kosteleck\'y V A, Lane C D and Okamoto T 2001 \textit{Phys. Rev. Lett.} \textbf{87} 141601; Amelino-Camelia G, Ellis J R, Mavromatos N E, Nanopoulos D V and Sarkar S 1998 \textit{Nature} \textbf{393} 763; Magueijo J and Smolin L 2003 \textit{Phys. Rev} D \textbf{67} 044017

\bibitem{String2} Kosteleck\'y V A and Samuel S 1989 \textit{Phys. Rev} D \textbf{39} 683; Kosteleck\'y V A and Potting R 1991 \textit{Nucl. Phys.} B \textbf{359} 545

\bibitem{SME} Colladay D and Kosteleck\'y V A 1997 \textit{Phys. Rev} D \textbf{55} 6760; Colladay D and Kosteleck\'y V A 1998 \textit{Phys. Rev} D \textbf{58} 116002; Kosteleck\'y V A 2004 \textit{Phys. Rev} D \textbf{69} 105009

\bibitem{Nambu} Nambu Y 1968 \textit{Suppl. Prog. Theor. Phys.} \textbf{E68} 190 (1968)

\bibitem{Abelian} Escobar C A and Urrutia L F 2015 \textit{Phys. Rev} D \textbf{92} 025042; Franca O J, Montemayor R and Urrutia L F 2012 \textit{Phys. Rev} D \textbf{85} 085008; Azatov A T and Chkareuli J L 2006 \textit{Phys. Rev} D \textbf{73} 065026

\bibitem{YM2} Chkareuli J L and Kepuladze Z R 2007 \textit{Phys. Lett.} B \textbf{644} 212; Chkareuli J L and Jejelava J G 2008 \textit{Phys. Lett.} B \textbf{659} 754; Chkareuli J L,  Froggatt C D and Nielsen H B 2009 \textit{Nucl. Phys.} B \textbf{821} 65; Chkareuli J L, Froggatt C D, Jejelava J G and Nielsen H B 2008 \textit{Nucl. Phys.} B \textbf{796} 211

\bibitem{Grav} Chkareuli J L, Jejelava J G, and Tatishvili G 2011 \textit{Phys. Lett.} B \textbf{696} 124

\bibitem{Conditions} Escobar C A and Urrutia L F 2015 \textit{Phys. Rev} D \textbf{92} 025013



\end{thebibliography}
\end{document}